\pdfoutput=1

\documentclass[11pt]{article}
\usepackage{amsmath}
\usepackage{multirow}
\usepackage{amsfonts}
\usepackage{algorithm}
\usepackage{algpseudocode}
\usepackage{siunitx}
\usepackage{tikz}
\usepackage{pgfplots}
\definecolor{LightPink}{rgb}{1.0, 0.75, 0.8}
\definecolor{Lavender}{rgb}{0.9, 0.9, 1.0}
\usepackage{graphicx}
\usepackage{pgfplots}  
\usepackage{booktabs}
\usepackage{listings} 
\usepackage{natbib}
\usepackage[final]{acl}
\usepackage{times}
\usepackage{latexsym}

\usepackage[T1]{fontenc}

\usepackage[utf8]{inputenc}

\usepackage{microtype}

\usepackage{inconsolata}

\usepackage{graphicx}

\newcommand\blfootnote[1]{%
  \begingroup
  \renewcommand\thefootnote{}\footnote{#1}%
  \addtocounter{footnote}{-1}%
  \endgroup
}

\title{Learning from Few Samples: A Novel Approach for High-Quality Malcode Generation}

\author{Haijian Ma\textsuperscript{1*}, \ Daizong Liu\textsuperscript{2*}, \ Xiaowen Cai\textsuperscript{1*}, \ Pan Zhou\textsuperscript{1$\dagger$},  \ Yulai Xie\textsuperscript{1$\dagger$}, \\
  \textsuperscript{1}Huazhong University of Science and Technology, Wuhan, China \\
  \textsuperscript{2}Peking University‌, Beijing, China  \\
  \texttt{\{mhj,xwcai,panzhou,ylxie\}@hust.edu.cn} \\
  \texttt{dzliu@stu.pku.edu.cn}\\
  } 

\begin{document}
\maketitle
\begin{abstract}
Intrusion Detection Systems (IDS) play a crucial role in network security defense. However, a significant challenge for IDS in training detection models is the shortage of adequately labeled malicious samples.
To address these issues, this paper introduces a novel semi-supervised framework \textbf{GANGRL-LLM}, which integrates Generative Adversarial Networks (GANs) with Large Language Models (LLMs) to enhance malicious code generation and SQL Injection (SQLi) detection capabilities in few-sample learning scenarios.
Specifically, our framework adopts a collaborative training paradigm where: (1) the GAN-based discriminator improves malicious pattern recognition through adversarial learning with generated samples and limited real samples; and (2) the LLM-based generator refines the quality of malicious code synthesis using reward signals from the discriminator. 
The experimental results demonstrate that even with a limited number of labeled samples, our training framework is highly effective in enhancing both malicious code generation and detection capabilities.
This dual enhancement capability offers a promising solution for developing adaptive defense systems capable of countering evolving cyber threats.
\end{abstract}

\blfootnote{
\textsuperscript{$*$}Equal contributions. ~~~~\textsuperscript{$\dagger$}Corresponding authors.}

\section{Introduction}
\label{sec:introduction}
The rapid evolution of cyber threats has exposed the limitations of traditional intrusion detection systems (IDS) in defending against modern attacks. A key challenge lies in the scarcity and lack of diversity in malicious samples—often referred to as black samples—which are essential for training robust detection models \citep{kavitha2024machine} and designing effective honeypots \citep{bp2024literature}.

\begin{figure}[t!]
  \centering
  \includegraphics[width=\columnwidth]{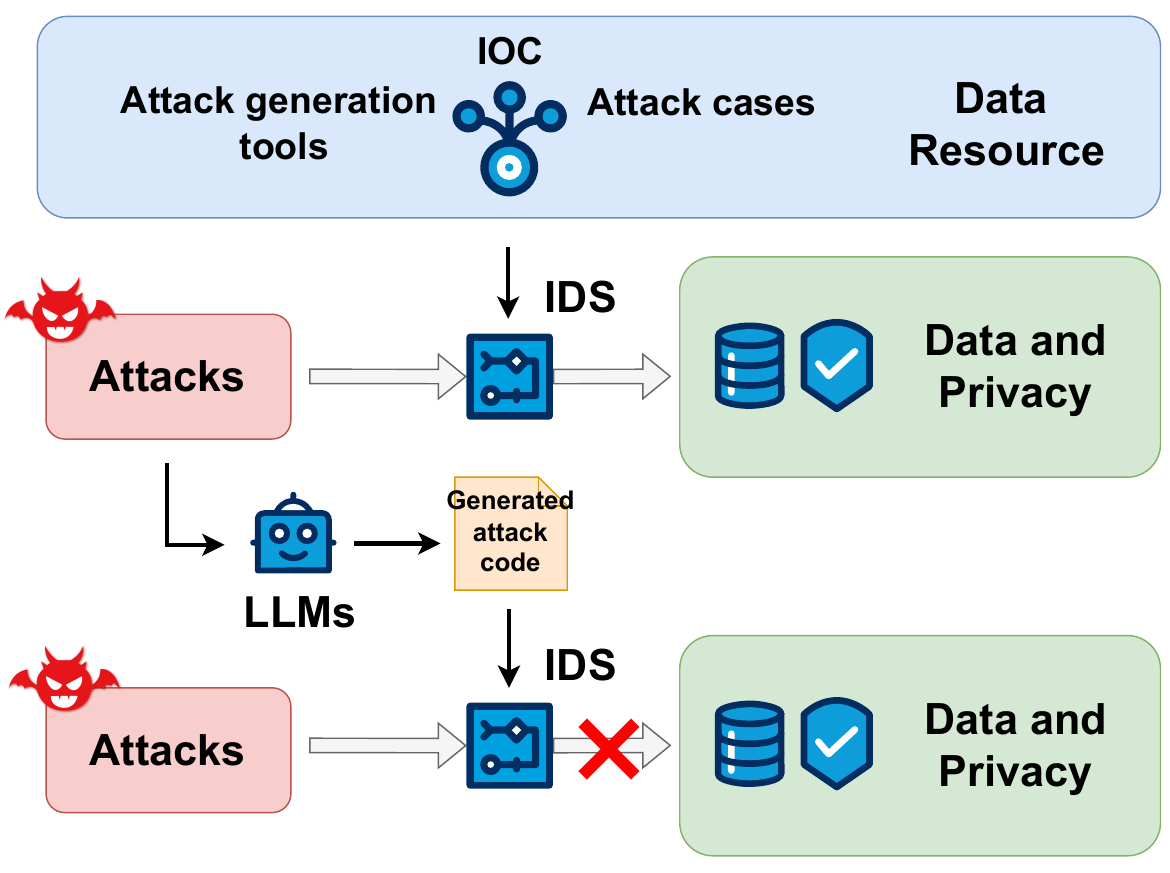}
  \vspace{-10pt}
  \caption{Illustration of our motivation.}
  \vspace{-10pt}
  \label{figure:teaser}
\end{figure}

Most IDSs today rely on three types of black sample sources: real-world attack data, open-source malcode generators, and threat intelligence-based Indicators of Compromise (IOC). However, each has notable drawbacks. Real-world attack cases are often limited due to privacy issues, legal constraints, and organizations’ reluctance to disclose breaches \citep{lu2022gan}, which reduces their usefulness for training generalized models. IOC data, while informative, is inherently reactive and typically lags behind emerging threats \citep{asiri2023understanding}, especially when facing increasingly diverse attack patterns \citep{izuazu2025explainable}. Furthermore, attackers often use obfuscation techniques to evade detection, undermining the reliability of IOC-based approaches. Therefore, a good malcode generator like the one in Figure~\ref{figure:teaser} is what we desperately need to improve the defenses of IDS. 

Large language models (LLMs) have demonstrated strong code generation capabilities \citep{jiang2024survey}. However, research has shown that their performance can degrade when trained on low-quality or insufficient datasets \citep{gu2024effectiveness}. While existing open-source attack generation tools provide synthetic samples, they often lack the complexity and variability required to simulate realistic attack scenarios. The limitations stem from both the generative capacity of the underlying models \citep{xu2023sql} and the quality and quantity of the training data \citep{setiyaji2024technique}. Therefore, this motivates us \textbf{to propose an LLM-driven approach that can generate high-quality malicious samples using only a small number of labeled instances, thereby enhancing the overall detection capability of intrusion detection systems (IDS).}

To address these limitations, we introduce GANGRL-LLM, a novel semi-supervised framework that synergistically integrates the complementary strengths of Generative Adversarial Networks (GANs) and Large Language Models (LLMs). Unlike conventional approaches, our framework leverages the powerful code-generation capabilities of large language models (LLMs), while incorporating a two-tiered GAN-based structure augmented with reinforcement learning reward signals to effectively guide the iterative training process.

A key innovation lies in the use of the discriminator’s output as a dynamic reward signal in the GAN, which actively guides the generator toward producing increasingly sophisticated and attack-like code structures. To further enhance training stability and prevent mode collapse, we incorporate contrastive constraints that preserve semantic consistency with real-world samples. The resulting two-tiered GAN-based structure enables collaborative, bidirectional learning between the generator and discriminator, significantly boosting malcode generation performance even under extreme data-scarcity conditions.

Our main contributions are summarized as follows:
\begin{itemize}
\item We introduce a novel multi-model collaborative learning framework for malcode generation which is different from text generation (more details in \ref{Deeper Discussion}), enabling joint training of generator and discriminator components to improve the robustness and generalization of detection models.
\item Our framework combines a two-tiered GAN-like architecture with LLMs, where the generator is trained with guidance from the discriminator’s output used as a reward signal. This strategy effectively enhances training stability and sample generation quality, especially when labeled data is limited.
\item Experimental results demonstrate that our framework not only effectively improves the generation quality of malicious samples and the detection accuracy of IDS models in few-shot learning settings but also exhibits transferability. 
\end{itemize}

\begin{figure*}[!t]
  \centering
  \includegraphics[trim={30 185 30 180}, clip,width=\textwidth]{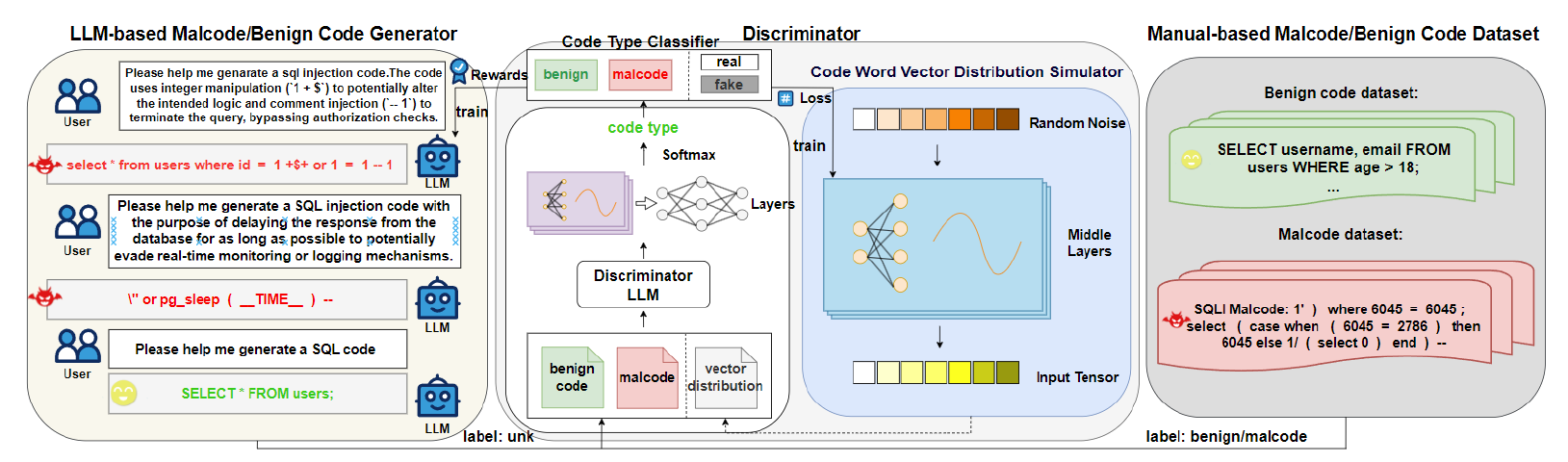}
  \caption{Framework overview of our proposed GANGRL-LLM.}
  \label{figure1:framework}
\end{figure*}

\section{Related Work}

\subsection{Text/Code Generation}
Recent advancements in code generation have leveraged pre-trained language models, such as GPT-3 and Codex, to generate high-quality code from natural language descriptions. Studies \citep{chen2021evaluating} have shown that large-scale pre-training on diverse datasets can lead to impressive performance on a range of code generation tasks. Further research has emphasized the importance of fine-tuning on high-quality and diverse datasets to improve the model's performance. A small or non-diverse dataset is prone to overfitting, where the model performs well on the training data but poorly on unseen data \citep{yin2017syntactic}.  Experimental results show that when the fine-tuning dataset is insufficient, the performance of LLMs in domain-specific code generation significantly declines. For instance, ChatGPT's average CodeBLEU score drops by 51.48\% in domain-specific tasks \citep{gu2024effectiveness}. This decline is primarily attributed to the model's unfamiliarity with using specific domain libraries.

Due to the performance degradation of fine-tuned models in code generation under few-shot labeling conditions, we employ token-level adversarial rewards to guide the model's learning and optimization. By leveraging the discriminator's judgments during the collaborative training of multiple models, we effectively direct the code generation process of the generative model.

\subsection{GAN for Text Generation}
Generative Adversarial Networks (GANs) \citep{goodfellow2020generative} were originally designed to generate realistic data samples and are particularly effective in scenarios with limited labeled data. However, one of the major challenges in applying GANs to text generation is their difficulty in handling discrete data. Unlike continuous outputs typical of GAN generators, textual data consists of discrete symbols such as words or characters, making gradient-based optimization through backpropagation problematic \citep{de2021survey}. 
Generative Adversarial Networks (GANs) were initially used for image generation.While \textbf{SeqGAN} \cite{yu2017seqgan} \textbf{LeakGAN} \cite{guo2018long} and \textbf{MaliGAN} \cite{che2017maximum} have made significant strides in sequence generation tasks, they still have notable limitations. The reward signals in existing methods are often sparse, which leads to slow convergence of the generator and unstable output quality. During training, particularly when generating diverse content, it is difficult to ensure convergence, leading to lower quality in the generated text or code. So we propose a novel design based on a discriminator reward mechanism. Unlike traditional cross-entropy loss, our model adjusts the generator's loss by using real-time rewards derived from the discriminator's output probability which is more effective and stable.

\subsection{Security Model}
Network attacks \cite{dong2025stabilizing,dong2025improving,dong2024robust,dong2024adversarially,dong2024advdistill,dong2023enemy,dong2022improving,dong2023restricted,dong2024survey,liu2025seeing,liu2022imperceptible,liu2023point,liu2023robust,liu2024explicitly,liu2025imperceptible,liu2024a,liu2024pandora,hu2022exploring,tao20233dhacker,yang2024hiding,cai2024frequency,cai2025imperceptible,cai2025imperceptibletransfer} such as SQL injection (SQLi) remains a significant security threat to web applications. Traditional detection methods, such as static and dynamic analysis, face challenges with dynamically generated queries and often miss certain attack vectors.

Recent advancements have seen the rise of machine learning-based detection methods. For instance, models like Random Forest \citep{joshi2014sql} and SVM \citep{zulu2024enhancing} have been applied for classifying malicious SQL queries. However, these methods still require handcrafted features and struggle with detecting more sophisticated attacks. Deep learning models such as CNNs \citep{luo2019cnn} and RNNs \citep{fang2018wovsqli} have shown promise in learning representations of SQL queries and detecting more complex injection attempts. While these models are effective, they still require large labeled datasets and considerable computational resources.

Despite these advancements \cite{fang2025your,fang2023hierarchical,fang2022multi,fang2023you,fang2025adaptive,fang2024fewer,fang2024multi,fang2024not,fang2025turing,fang2024rethinking,fang2023annotations,fang2021unbalanced,fang2025adaptive,fang2020v,fang2021animc,fang2024uncertainty,fang2020double,liu2021context,liu2020jointly}, challenges such as sparse labeled data and model stability remain. To address these issues, our SQL security model uses a discriminator reward mechanism, leveraging reinforcement learning and real-time reward signals to improve detection performance even with limited labeled data, enhancing both accuracy and efficiency.

\section{Methodology}
\subsection{Overall Framework}
In certain security domains, the availability of labeled malicious samples for model training is severely limited. Consequently, we aim to design a training framework that can guide large models to learn and expand their generation capabilities for generating high-quality malcode, even when only a small number of labeled black samples are available.

As shown in Figure~\ref{figure1:framework}, our adversarial generation framework combines a code generator and a discriminator in a reinforcement learning loop. The generator produces SQL injection (SQLi) code snippets based on the SQL injection methods described in the prompt, while the discriminator distinguishes between malicious (SQLi) and benign samples. The training proceeds iteratively for several rounds, with two phases per round:
1) Generator Optimization: The generator uses cross-entropy and a reward signal (SQLi probability) predicted by the discriminator as to train.
2) Discriminator Training: Generated SQLi samples (labeled as $Unk$) from the code generator and Manual-based Code Dataset are used to update the discriminator.

\subsection{Discriminator}
The integration of Generative Adversarial Networks (GAN) with the BERT model \citep{croce2020gan} enables the discriminator to achieve more accurate judgments than a standard BERT model, even when only a small amount of labeled data and a certain quantity of unlabeled data are available. In this work, a code word vector distribution simulator and a classifier complement each other to enhance the discriminator's performance. The simulator continuously learns to produce the distribution of realistic code to deceive the classifier, while the classifier strives to distinguish between real data and the fake data simulated by the simulator, thereby improving its classification ability in the process. This can also provide a crucial foundation for the reward scoring of the code generator.

Therefore, we employ two multi-layer perceptrons (MLPs): one as the code word vector distribution simulator and another as the code type classifier. The simulator receives a random noise vector as input, which is transformed to the input tensor, simulating the distribution of the hidden states of real samples. To obtain the hidden states of the manual-based code samples and generated samples, we first encode them using the discriminator LLM. The generated noise, along with the manual-based code(labeled) and generated samples(unk), is then fed into the classifier. The classifier has two tasks: it classifies whether the sample is real or generated, and it also predicts the true class label of the sample.

\subsubsection{Code Type Classifier}
The classifier adopts the multi-layer perceptrons (MLPs) architecture to classify inputs into $k+1$ classes, where the first $k$ classes represent true data and the last class represents fake data. Among the first k classes, we further classify them into two categories: benign and SQL injection (SQLi) \citep{salimans2016improved}. Let $C$ and $S$ denote the classifier and simulator, with $p_c$ and $p_s$ representing the real data distribution and simulated distribution respectively. For semi-supervised learning with $k$-class classification, we extend the classifier objective as follows:

Define the classification probabilities:
$C(y|x, y{\in}\{1,...,k\})$ is the probability of example $x$ belonging to class $y$ and $C(y{=}k{+}1|x)$ is the probability of $x$ being generated (fake class)

The classifier loss combines supervised and unsupervised components:
\vspace{-.1cm}
\begin{equation}
    \mathcal{L}_C = \mathcal{L}_{C_{\text{sup}}} + \mathcal{L}_{C_{\text{unsup}}},
\end{equation}

\noindent where the supervised loss measures classification error:
\vspace{-.1cm}
\begin{equation}
    \mathcal{L}_C{_{\text{sup}}} = -\mathbb{E}_{(x,y)\sim p_c}\log\left[C(y|x, y{\in}\{1,...,k\})\right],
\end{equation}

\noindent and the unsupervised loss detects simulated examples:
\begin{equation}
\begin{split}
    \mathcal{L}_{C_{\text{unsup}}} = & -\mathbb{E}_{x \sim p_c} \log \left[1 - C(y=k+1|x)\right] \\
                                      & - \mathbb{E}_{x \sim p_s} \log \left[C(y=k+1|x)\right].
\end{split}
\end{equation}

$\mathcal{L}_{C_{\text{sup}}}$ penalizes misclassification of real examples among original $k$ classes.$\mathcal{L}_{C_{\text{unsup}}}$ contains two terms:
(1) The first term prevents misclassifying real examples as fake.
(2) The second term improves fake example detection.

\subsubsection{Code Word Vector Distribution Simulator}
For enhancing the diversity of the dataset put in the classifier for stronger detection capability, we have designed a code word vector distribution simulator to simulate the token distribution of real samples from random noise. Additionally, these fake samples can be used to train the classifier, improving its classification capabilities.

The simulator $S$ optimizes a composite loss function with adversarial training and feature matching regularization:
\vspace{-.3cm}
\begin{equation}
\begin{split}
    \mathcal{L}_S = & \underbrace{-\mathbb{E}_{x_s \sim p_s} \log(1 - C(y=k+1|x_s))}_{\text{Adversarial Loss}} \\
                    & + \lambda \underbrace{\mathbb{E}\left[\left\|\mu_{\text{real}} - \mu_{\text{fake}}\right\|_2^2\right]}_{\text{Feature Matching Loss}},
\end{split}
\end{equation}

where  $C(y{=}k{+}1|x_s)$ denotes the classifier's probability estimate that a simulated sample $x_s$ belongs to the fake class. $\mu_{\text{real}} = \mathbb{E}_{x_r\sim p_c}[f_c(x_r)]$ is the mean feature vector of real samples. $\mu_{\text{fake}} = \mathbb{E}_{x_s\sim p_s}[f_c(x_s)]$ is the mean feature vector of simulated samples.$f_c(\cdot)$ represents the classifier's penultimate layer features.

\subsection{Code Generator with LLM}
Large Language Models (LLMs) exhibit strong general-purpose code generation capabilities. To train an effective generator for malicious code, we leverage LLMs as an external code generation tool. However, since the quality and effectiveness of the generated code are uncertain, we label all such samples as "unk" during the training of the discriminator. By incorporating these unlabeled samples into the training process, our approach not only aligns with the adversarial design principles of GANs but also enhances the generator's ability to learn logical and structurally coherent code patterns through iterative refinement.

The generator is initialized from the pre-trained Qwen2.5Coder, a Transformer-based LLM specialized in code generation. Given a prompt x, it generates SQLi code y by sampling from the probability distribution:
\vspace{-.1cm}
\begin{equation}
y \sim P_\theta(y | x) = \prod_{t=1}^T P_\theta(y_t | y_{<t}, x),
\end{equation}
where $\theta$  denotes the generator’s parameters. 
\subsection{Code Generation Training Protocol}
To achieve joint optimization of the generative model and the discriminator, while considering the small differences in multiple sampling runs, the complexity of the process, and the high resource consumption during training, we use cross-entropy loss as an anchor to prevent model collapse. The reward term is based on the log probability (rather than the raw probability) that the discriminator assigns to the generated code being malicious, which results in smoother gradients. By combining these two elements into the loss function, we create a more dense feedback signal, guiding the code generation to be more purposeful and standardized.

\label{subsec:gen_train}
The Qwen generator undergoes policy optimization with adaptive reward shaping, as detailed in Algorithm~\ref{alg:gen_train}.

\begin{algorithm}[t]
\caption{Code Generator Training}
\label{alg:gen_train}
\begin{algorithmic}[1]
\For{each batch $\mathcal{B} \in \mathcal{D}_{\text{train}}$}
    \State Parse batch: $\{\mathbf{x}_i, \mathbf{y}_{i_{\text{gen}}}, \mathbf{y}_{i_{\text{real}}}\} \gets \mathcal{B}$
    
    \State Compute rewards:
    \State \quad Tokenize $\mathbf{y}_{\text{gen}}$: 
        \State \qquad $\mathbf{H}_{\text{gen}} \gets \text{Tokenizer}(\mathbf{y}_{\text{gen}})$
    \State \quad Get discriminator output: 
        \State \qquad $D(\mathbf{H}_{\text{gen}}) \to (\cdot, \mathbf{logits}, \mathbf{probs})$
    \State \quad Extract reward: 
        \State \qquad $r_i \gets \mathbf{probs}[0,1]$
        \Comment{Fake class probability}
    
    \State Prepare model inputs:
    \State \quad $\mathbf{H}_{\text{in}} \gets \text{FormatPrompt}(\mathbf{x}_i)$
    \Comment{Alpaca prompt template}
    \State \quad $\mathbf{H}_{\text{target}} \gets \text{Tokenize}(\mathbf{y}_{\text{real}})$
    
    \State Forward pass:
    \State \quad $\mathcal{L}_{\text{MLE}} \gets -\log p_{\theta}(\mathbf{y}_{\text{real}}|\mathbf{H}_{\text{in}})$
    \State \quad $\mathcal{L}_{\text{RL}} \gets -\lambda \log r_i$ \Comment{Reward-augmented loss}
    
    \State Update parameters:
    \State \quad $\theta \gets \theta - \eta \nabla_{\theta}(\mathcal{L}_{\text{MLE}} + \mathcal{L}_{\text{RL}})$
    \State \quad Apply gradient clipping: $\|\nabla\theta\|_2 \leq 1.0$
\EndFor
\end{algorithmic}
\end{algorithm}

\subsubsection{Adaptive Reward Weighting} 
We introduce dynamic decay to adjust the reward scores provided by the discriminator, preventing the training from becoming unstable in later stages due to an increase in the discriminator's capability, which could lead to excessively high reward scores. By doing so, we allow the generative model to rely more on the discriminator's feedback during the early stages of training, while gradually reducing this reliance in later stages to ensure greater stability and convergence of the model.
The mixing coefficient $\lambda$ decays exponentially during training:
\vspace{-.2cm}
\begin{equation}
    \lambda(t) = \alpha \times (\theta)^{t/T},
\end{equation}
where $\alpha$ is the weight of the reward value,$\theta$ is the reward adaptation coefficient per round,$t$ is the current epoch, and $T$ is the total number of training epochs.

\subsubsection{Loss Formulation}
By using the discriminator's rewards for unsupervised training and the generator's cross-entropy loss for supervised learning, co-training can proceed stably. This combined approach allows the generative model to quickly learn to generate high-quality malicious code in the early stages and gradually reduce its reliance on the discriminator's feedback in later stages, thereby ensuring the stability and convergence of the entire training process.
The discriminator provides real-time feedback through a reward signal:
\vspace{-.3cm}
\begin{equation}
    r(\mathbf{y}_{\text{gen}}) = D(y{=}1|\mathbf{y}_{\text{gen}}),
\end{equation}
\vspace{-.1cm}
where $D(y{=}1|\cdot)$ denotes the probability of generated samples being classified as malicious by GANBERT.
\vspace{-.3cm}
\begin{equation}
    \mathcal{L}_{\text{total}} = \underbrace{-\mathbb{E}[\log p_\theta(\mathbf{y}_{\text{real}}|\mathbf{x})]}_{\text{Supervised Loss}} + \lambda \underbrace{\mathbb{E}[-\log r(\mathbf{y}_{\text{gen}})]}_{\text{Policy Gradient Term}}.
\end{equation}
The composite training objective combines maximum likelihood estimation with policy gradient rewards.
\begin{table}[ht]
  \centering
  \small
  \begin{tabular}{l|p{0.3\textwidth}} 
    \toprule
    \textbf{Section} & \textbf{String} \\
    \midrule\midrule
    \texttt{Instruction} & Please help me genarate a sql injection code.The SQL injection uses a single-quote to break out of the original query, followed by a UNION ALL statement to combine results from another SELECT statement with all NULL values, and ends with a comment symbol `--` to ignore any remaining original query. \\
    \midrule
    \texttt{Input} & \verb|\| \\
    \midrule
    \texttt{Output} & \textbf{1\%'union all select null,null,null-- }\\
    \bottomrule
  \end{tabular}
  \caption{Example of training datasets adhering to the Alpaca training template.}
  \label{tab:alpaca}
\end{table}

\section{Experiments}

\subsection{Experimental Setup}

\textbf{Datasets}. We utilized a subset of the SQL Injection (SQLi) dataset \citep{kaggle_ml_sql_injection} for training, while reserving 100 instances as the generation test set. Based on the specific implementation methods of SQLi in the dataset, we designed corresponding prompts as instructions and used the associated SQLi code snippets as outputs. The training dataset adheres to the Alpaca training template, as illustrated in Table~\ref{tab:alpaca}. \\\textbf{Environment}.Our experiments were conducted on three NVIDIA RTX A5000 (24GB) GPUs, with an Intel(R) Xeon(R) Platinum 8222L CPU @ 3.00GHz. The system ran Ubuntu 22.04.5, and the experiments were implemented in Python 3.11.10. We used PyTorch version 2.5.1 for model training and evaluation, with CUDA 12.1 for GPU acceleration. \\\textbf{The Process and Related Parameters}. As for data preprocessing, we utilize the GPT-4 API for generating the unlabeled training samples of SQL code with its description. As for hyperparameters, both our discriminator and generator are trained with a learning rate of 1e-5 for 20 epochs, and we set the batch size to 64. The $\epsilon$ is 0.05 and $\theta$ is 0.9 in the formula of the adaptive weight $\lambda$. Our data is divided into four categories: benign, malicious, unlabeled, and fake. This setup is designed to optimize the performance of our model, ensuring high-quality malicious code generation even with limited initial data, thereby enhancing the robustness and effectiveness of security models trained on these augmented datasets. We will provide more detailed descriptions of the experimental setup and release our codes to ensure the reproducibility of our research in the revision.

\subsection{The Validity of Code Generation}
In this section of the experiment, we will present a comparative analysis of three versions of the Qwen2.5coder model(1.5B)\citep{hui2024qwen2}: no fine-tuning, finetuned, finetuned and trained by GANGRL-LLM.

All models will be evaluated on the same set of 100 prompts designed to generate SQL injection code corresponding to given requirements.
For each generated result, we will employ the Qwen2.5Turbo \citep{yang2024qwen2} using API to evaluation system to score the outputs on a scale from 1 to 10 based on the criteria: adherence to the prompt, complexity of the code, effectiveness of the sql injection and correctness of the code(more details in \ref{Scoring Basis}). This system leverages the advanced capabilities of Qwen2.5Turbo to provide detailed assessments across multiple dimensions.
\begin{table}[h]
\centering
\small
\begin{tabular}{c|c|c}
\toprule
\textbf{Number of datasets} & \textbf{Model} & \textbf{Score} \\\midrule\midrule
\multirow{2}*{no fine-tuning} & Qwen2.5coder & 5.58 \\
& GANGRL-LLM & \textbf{5.64$\uparrow$} \\ \hline
\multirow{2}*{1000} & Qwen2.5coder & 5.275 \\
& GANGRL-LLM & \textbf{5.74$\uparrow$}  \\ \hline
\multirow{2}*{2000} & Qwen2.5coder & 6.35 \\
& GANGRL-LLM & \textbf{6.4$\uparrow$}  \\\bottomrule
\end{tabular}
\caption{Scores(ranging from 1 to 10) for different models with varying numbers of fine-tuning datasets.}
\label{tab:scores}
\end{table}

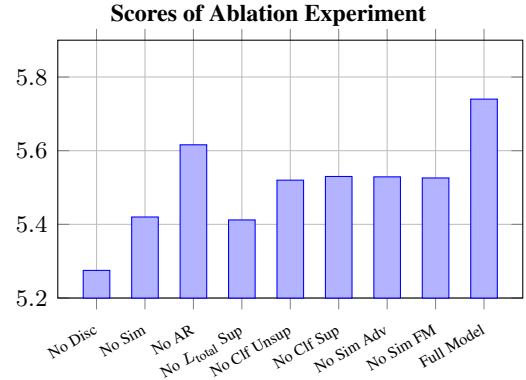
\begin{figure}[htbp]
    \centering
    \textbf{\small Scores of Ablation Experiment} 
    \vspace{10pt} 
    \begin{tikzpicture}
        \begin{axis}[
            width=\linewidth,
            height=5cm,
            ybar,
            bar width=0.35cm,
            symbolic x coords={
                No Disc,
                No Sim,
                No AR,
                No $L_{\text{total}}$ Sup,
                No Clf Unsup,
                No Clf Sup,
                No Sim Adv,
                No Sim FM,
                Full Model
            },
            xtick=data,
            x tick label style={
                rotate=30,
                anchor=north east,
                font=\tiny
            },
            ylabel={}, 
            ymin=5.2, ymax=5.9,
            ytick={5.2,5.4,5.6,5.8},  
            yticklabel style={font=\footnotesize},  
            grid=major,
            axis background/.style={fill=white},
            every axis plot/.append style={fill=blue!60}
        ]
        \addplot+[] coordinates {
            (No Disc, 5.275)
            (No Sim, 5.42)
            (No AR, 5.616)
            (No $L_{\text{total}}$ Sup, 5.412)
            (No Clf Unsup, 5.52)
            (No Clf Sup, 5.53)
            (No Sim Adv, 5.529)
            (No Sim FM, 5.526)
            (Full Model, 5.74)
        };
        \end{axis}
    \end{tikzpicture}
    \vspace{-15pt} 
    \caption{GANGRL-LLM Ablation Study. Abbreviations: No Disc = without discriminator; No Sim = No simulator; No AR = No adaptive reward weight; No $L_{\text{total}}$ Sup = No $L_{\text{total}}$'s supervised loss; No Clf Unsup = No classifier unsupervised loss; No Clf Sup = No classifier supervised loss; No Sim Adv = No simulator adversarial loss; No Sim FM = No simulator feature matching loss; Full Model = GANGRL-LLM (full model)}
    \label{figure:ablation}
\end{figure}

\begin{figure*}[htbp]
    \centering
    \includegraphics[width=1\textwidth]{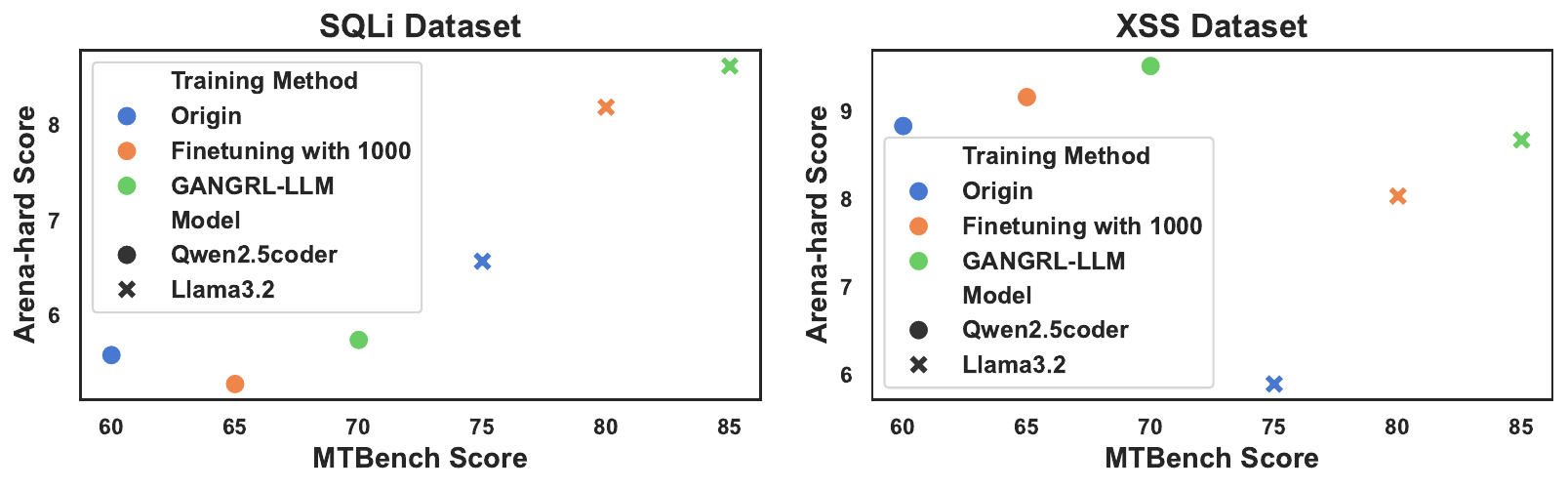} 
    \caption{Scores for different models trained on the SQLi and XSS dataset using various training methods(no operation, finetuned with 1000 samples, trained with GANGRL-LLM).} 
    \label{fig:m} 
\end{figure*}

\definecolor{custom_green}{rgb}{0.2, 0.8, 0.2} 
\definecolor{custom_red}{rgb}{0.8, 0.2, 0.2} 
\begin{table*}[ht]
\small
\centering

\begin{tabular}{l|l|l|l|l|l}
\toprule
\textbf{Operation} & \textbf{MODEL NAME} & \textbf{Accuracy} & \textbf{Precision} & \textbf{Recall} & \textbf{F1} \\ \midrule\midrule
\multirow{5}{*}{no operation} & CNN          & 0.904 & 0.9985 & 0.8813  & 0.9362 \\ 
                              & Naive Bayes  & 0.828 & 0.8243 & 0.9975   & 0.9027 \\ 
                              & SVM          & 0.489 & 1 & 0.3613   & 0.5308 \\ 
                              & KNN         & 0.807 & 0.8056 & 0.8975   & 0.8924 \\  
                              & Decision Tree & 0.917 & 0.9986 & 0.8975   & 0.9454 \\ \midrule
\multirow{5}{*}{Switch 1000 training datasets} & CNN & \textbf{0.91}  & \textbf{1}                   & \textbf{0.8875}   & \textbf{0.9404} \\ 
                                               & Naive Bayes  & \textbf{0.932} & \textbf{0.9973} & 0.9175& \textbf{0.9557} \\ 
                                               & SVM          & \textbf{0.498} & 1                   & \textbf{0.3725}   & \textbf{0.5428}  \\ 
                                               & KNN          & 0.807 & 0.8056 & 1  & 0.8924 \\ 
                                               & Decision Tree & \textbf{0.921} & \textbf{1}                   & \textbf{0.9013}  & \textbf{0.9481}  \\ \midrule
\multirow{5}{*}{Add 1000 training datasets}   & CNN          &\textbf{0.913} & \textbf{1} & \textbf{0.8913}   & \textbf{0.9425} \\  
                                               & Naive Bayes  & \textbf{0.932} & \textbf{0.9972} & 0.9175  & \textbf{0.9557} \\  
                                               & SVM          & \textbf{0.8} & 0.8 & \textbf{1}& \textbf{0.8889} \\ 
                                               & KNN          & 0.807 & 0.8056 & 1  & 0.8924 \\ 
                                               & Decision Tree & \textbf{0.919} & \textbf{1}                   & \textbf{0.8988}  & \textbf{0.9467} \\ \midrule
\multirow{5}{*}{Add 2000 training datasets}   & CNN          & \textbf{0.92}  & \textbf{0.9986} & \textbf{0.9013}  &\textbf{0.9474} \\ 
                                               & Naive Bayes  & \textbf{0.93}  & \textbf{0.9946}  & 0.9175   & \textbf{0.9545} \\ 
                                               & SVM          & \textbf{0.8} & 0.8 & \textbf{1}& \textbf{0.8889} \\ 
                                               & KNN          & 0.807 & 0.8056 & 1        & 0.8924 \\ 
                                               & Decision Tree & \textbf{0.92}  & \textbf{1}                   & \textbf{0.8988}  & \textbf{0.9475} \\ \bottomrule
\end{tabular}

\caption{Models performance for different training Qwen-generated set composition. Bold numbers indicate an increase.}
\label{tab:ml}
\end{table*}
\begin{table*}[h]
\centering
\small
\begin{tabular}{@{}l|l|c|c|cc@{}}
\toprule
\textbf{Data} & \textbf{Accuracy} & \textbf{F1} & \textbf{Precision} & \textbf{Recall} \\ \midrule\midrule
train: 600, test: 400 & 0.75 & 0.857143 & 0.75 & 1 \\\midrule
train: 600, test: 400 (Switch 200) & \textbf{0.865} &\textbf{0.917431} & \textbf{0.847458} & \textbf{1} \\\midrule
train: 800, test: 400 (Add 200) & \textbf{0.9225} & \textbf{0.950715} & \textbf{0.908815} & \textbf{0.996667} \\\midrule
train: 900, test: 400 (Add 300) & \textbf{0.9575} & \textbf{0.972447} & \textbf{0.946372} & \textbf{1} \\ \bottomrule
\end{tabular}
\caption{Performance metrics for different data set composition. Bold numbers indicate an increase.}
\label{tab:performance_metrics}
\end{table*}
\vspace{-10pt}

As shown in Table~\ref{tab:scores}, models trained with the GANGRL-LLM framework demonstrate improved generation performance compared to their original versions. We also experimented with unsupervised learning using policy gradients from reinforcement learning (Table~\ref{tab:scoresrl}), but the results were less effective than our proposed method. Notably, GANGRL-LLM helps maintain model performance even when training data is limited. As more data becomes available, model generalization improves, but the relative gain from our framework decreases. This indicates that GANGRL-LLM is especially beneficial for enhancing generation capabilities in low-data regimes.
\begin{table}[h]
\centering
\small
\begin{tabular}{l|c}
\toprule
\textbf{Model} &  \textbf{Recall (\%)} \\
\midrule\midrule
Gamma-TF-IDF &  99.2\\
I-TF-IDF &  60.9\\
EP-CNN &  98.0\\
SQL-MLP &  98.2\\
SQL-LSTM &  95.7\\
ASTNN &  99.2\\
Trident &  99.4\\
\textbf{Our discriminator} & \textbf{99.9$\uparrow$}\\\bottomrule
\end{tabular}
\caption{Reference benchmarks and their precision and recall scores on the SQLi dataset.}
\label{tab:benchmarks}
\end{table}
We also utilized the AI SQLi detection system from Chaitin Tech \citep{chaitin} to test the data generated using our framework. Our findings indicate that out of 1,000 generated samples, 997 were successfully detected as SQLi code. Under a dataset of 1,000 labeled samples, the effectiveness rate of our generated samples is 99.7\%.
\subsection{Ablation Experiment}
As shown in Figure~\ref{figure:ablation}, we perform ablation studies by removing different components of GANGRL-LLM using 1000 training samples. The full model achieves the highest score of 5.74, while all variants exhibit varying degrees of performance degradation. Removing the discriminator leads to the most significant drop, highlighting the importance of adversarial learning. Components such as the code word vector distribution simulator, classifier supervised loss and $L_{\text{total}}$'s supervised loss also play notable roles in maintaining performance. Overall, these results highlight the roles of different components in guiding generator learning and maintaining model effectiveness under limited data conditions.

\subsection{Migratability}
We employed two models, Llama3.2\cite{grattafiori2024llama} and Qwen2.5coder, to train on datasets pertaining to SQL Injection (SQLi) and Cross-Site Scripting (XSS) attacks. As illustrated in the Figure~\ref{fig:m}, our GANGRL-LLM approach demonstrates remarkable transferability across both models and datasets, even when trained with a limited number of samples. This indicates that our method effectively enhances the capability of different models to generate malicious code across various datasets.
\subsection{The Validity of Generated Dataset}
We utilized the publicly available SQLi dataset \citep{kaggle_sql_injection_dataset} and generated additional data using the Qwen2.5coder model, which was fine-tuned on 1,000 samples and subsequently trained using the GANGRL-LLM framework.
\begin{table}[h]
\centering
\small
\begin{tabular}{l|c|c|c}  
\toprule
\textbf{Model} & \textbf{GF} & \textbf{DM} & \textbf{Score} \\
\midrule\midrule
\multirow{4}{*}{Llama3.2} 
    & RL       & BERT with GAN       & 4.28 \\
    & GAN      & Codex + RLHF BERT  & 4.55 \\
    & GAN      & BERT with MixMatch & 4.60 \\
    & \textbf{GAN}      & \textbf{BERT with GAN}      & \textbf{5.23} \\
\midrule  
\multirow{4}{*}{Qwen2.5coder} 
    & RL       & BERT with GAN       & 5.35 \\
    & GAN      & Codex + RLHF BERT  & 5.44 \\
    & GAN      & BERT with MixMatch & 5.46 \\
    & \textbf{GAN}     & \textbf{BERT with GAN}      & \textbf{5.74} \\
\bottomrule
\end{tabular}
\caption{Comparison of scores for different model series, generation frameworks, detection methods. GF means Generation Frameworks and DM means Detection Methods.}
\label{tab:method-comparison}
\end{table}

\begin{table*}[h]
\centering
\small
\begin{tabular}{p{0.25\textwidth}|p{0.65\textwidth}}
\toprule
\multicolumn{2}{c}{\textbf{Prompt}} \\
\midrule\midrule
\multicolumn{2}{p{0.9\textwidth}}{Please help me generate a SQL injection code. The SQL injection vulnerability in the provided code utilizes the \colorbox{LightPink}{regexp\_substring} function, exploits the \colorbox{LightPink}{char} function for potentially crafting a payload, and depends on improper handling of \colorbox{LightPink}{NULL values and logical conditions} to mask the malicious input.} \\
\toprule
\textbf{Model} & \textbf{Generated Code}\\
\midrule\midrule
Testset & call \colorbox{LightPink}{regexp\_substring}( \colorbox{Lavender}{repeat}(right(char(3702),0),500000000),\colorbox{LightPink}{null} ) and ( 9646 = 9646 \\ 
\midrule
Finetuned(1000 datasets) model & 1" ) ) as ywv where 5252 = 5252 and char(105) ||char(110) ||char(103) ||char(115) = \colorbox{LightPink}{regexp\_substring}(char(73) ||char(113) ||char(117) ||char(108),\colorbox{LightPink}{null}) -- \\
\midrule
Finetuned model and trained by GANGRL-LLM  & 1" ) ) and char(113) ||char(113) ||char(103) ||char(113) = \colorbox{LightPink}{regexp\_substring}(char(65) ||char(69) ||char(83),\colorbox{LightPink}{null}) and (( \colorbox{Lavender}{"zjzq" = "zjzq"} \\ 
\bottomrule
\end{tabular}
\caption{Examples of generated SQL code from different models.}
\label{tab:generated-code-examples-adjusted}
\end{table*}
From the results in Table~\ref{tab:ml}, Table~\ref{tab:ml2} and Table~\ref{tab:ml3}, it is evident that models trained using the GANGRL-LLM framework exhibit enhanced detection capabilities for SQLi code, even when fine-tuned with a limited number of samples. Initially, we replaced 1,000 malicious samples in the original training set with 1,000 samples generated by our model. The results demonstrated that most models showed improved detection performance. Building on this, we further expanded the training set by adding 1,000 and then 2,000 additional samples, all generated by our trained model. These augmented datasets consistently enhanced the performance of the models.

As shown in Table~\ref{tab:performance_metrics}, due to the superior capabilities of large models compared to traditional deep learning and machine learning methods, we further reduced the dataset size to 600 samples to simulate a scenario with limited training data. We utilized the bert-uncased model (specifically the uncased\_L-12\_H-768\_A-12 variant) \citep{devlin2018bert}for classification tasks. In our experiment, we replaced 200 samples in the original training set with generated samples. Subsequently, we conducted additional experiments to expand the dataset. The results indicate that incorporating these generated samples significantly enhances the effectiveness of model training.

We tested our discriminator, which was fine-tuned with 1,000 data samples from SQLi \citep{kaggle_sql_injection_dataset}  and subsequently trained using the GANGRL-LLM framework. The reference benchmarks used in our evaluation are listed in Table~\ref{tab:benchmarks}. These models include Gamma-TF-IDF \citep{Guan2022TFIDF}, I-TF-IDF \citep{li2020sql}, EP-CNN \citep{xie2019sql}, SQL-MLP \citep{tang2020detection}, SQL-LSTM \citep{tang2020detection}, ASTNN \citep{zhang2019novel}, and Trident \citep{li2024trident}. And in Figure~\ref{figure:accuracy}, the accuracy of our detector improves when training process remains stable.

As shown in Table~\ref{tab:benchmarks}, our discriminator achieved a notably high recall, indicating a strong ability to detect true positive instances. Comparative analysis shows that, despite using fewer training samples than those used in benchmark tests, our trained discriminator still achieves excellent recall performance. This demonstrates the effectiveness and resource efficiency of our framework, making it well-suited for cybersecurity applications where reliable detection models can be developed using only a limited number of training samples.

\subsection{Method Comparison}
In this section, we compare our method with some typical attack-generation frameworks and semi-supervised detection methods. In Table~\ref{tab:method-comparison}, we apply our framework to two open-source models: Llama3.2 and Qwen2.5-Coder, and modify the core components of the framework by incorporating a reinforcement learning (RL)\citep{sutton1998reinforcement} training mechanism, a Codex\citep{chen2021evaluating} + RLHF-based\citep{stiennon2020learning} BERT detector, and a MixMatch\citep{berthelot2019mixmatch} semi-supervised detection strategy, respectively. We conduct experiments using 1,000 training samples under few-shot settings. The experimental results show that our proposed GAN-based framework achieves much better performance than all tested baseline models under the same data constraints, demonstrating the effectiveness and transferability of our method. 
\subsection{Generation Results}
In Table~\ref{tab:generated-code-examples-adjusted}, in the first SQL injection code where char(3702) might produce an invisible or invalid character. This makes the attack more likely to be blocked by protective systems. It lacks complex conditions or logic bypass mechanisms, making it relatively straightforward but potentially ineffective against robust defenses.The segment of the third SQL injection code is relatively simple compared to the second one. It lacks advanced logical conditions or sophisticated evasion techniques.

In terms of conformity to the prompt and complexity and effectiveness of the SQL injection, the second SQL injection code is the most aligned and sophisticated. It closely matches the prompt(more details in appendix~\ref{malcode analysis}), making the injection harder to detect and more likely to bypass security measures. This indicates that our training framework can produce higher-quality code generated by the models.

\section{Conclusion}
Our work introduces GANGRL-LLM, that integrates GAN-like structure with LLMs in a multi-model collaborative training paradigm. By leveraging the discriminator’s output as a reward signal, our approach effectively guides the generator to produce high-quality malicious code even under data-scarce conditions, enhancing the generation capability of the model and improving the detection performance of IDS.

Experimental results demonstrate that our framework achieves strong performance in both SQL injection (SQLi) code generation and attack detection under few-shot learning settings. Moreover, The framework exhibits strong transferability, being adaptable to different datasets and model architectures addressing the challenges of limited malicious sample data in modern cybersecurity defense.


\section*{Limitations}
Despite the fact that GANGRL-LLM has been designed and shown to enhance the capabilities of detection models, there is still room for improvement, particularly in developing a more effective reward mechanism for the generative model. Additionally, the framework needs to be extended and optimized for application across multiple domains to increase its versatility.

One promising direction is to integrate malicious code samples from various security domains to train the LLM on a broader range of threats using limited samples from each domain. This approach could result in a more versatile model capable of providing comprehensive support for black-box testing and enhancing security detection models across multiple areas. Such a model would be invaluable for security researchers aiming to optimize their detection systems with diverse and high-quality training data.

\section*{Ethical Statement}
The SQL injection (SQLi) training datasets used in our experiments are sourced from open-source datasets. The SQLi code generated during the experiments has not been publicly released. The primary objective of this experiment is to enhance the model's capability to generate malicious code, thereby creating a higher-quality dataset of black samples for security research purposes.
\bibliography{custom}

\appendix
\section{Malcode}
In this work, we use sqli as a proxy for malicious code. SQL Injection (SQLi) is a code injection technique where attackers insert malicious SQL statements into an application's input fields to execute them on the database server. This exploits vulnerabilities in applications that improperly filter user inputs or lack stringent validation, allowing attackers to bypass authentication mechanisms and manipulate databases. Successful exploitation can lead to unauthorized access, data modification, or deletion, posing significant security risks.

\textbf{Error-Based SQL Injection}: In this type, attackers leverage error messages generated by the database to deduce its structure or extract sensitive data. For instance, submitting a username like \verb|' OR 1=1 --| and any password in a login form can result in executing:

SELECT * FROM users WHERE username = '' OR 1=1 --' AND password = 'anything';

This bypasses authentication and provides access if the first record corresponds to an admin account. By analyzing returned errors or successful logins, attackers gather insights into the database schema.

\textbf{Boolean-Based Blind SQL Injection:} When applications return only true or false responses, attackers can infer information by observing these responses. An example query might be:

' AND (SELECT COUNT(*) FROM users) > 0 AND '1'='1

If the page loads normally, it indicates the existence of the \texttt{users} table. This method is less efficient but effective for probing applications that do not display detailed error messages.

\textbf{Time-Based Blind SQL Injection:} In cases where no feedback is visible, attackers use time delays to infer data. For example, to check if the first character of the current database name is `a':

' AND IF(ASCII(SUBSTRING((SELECT DATABASE()),1,1))=97,SLEEP(5),null) AND '1'='1

A delay exceeding five seconds suggests the first character matches `a'. This technique is useful against applications protected by strict firewalls or intrusion detection systems.

\textbf{Union-Based SQL Injection:} Attackers can append additional SELECT queries using UNION operators to extract data from different tables. For instance:

SELECT * FROM products WHERE id = '1' UNION SELECT null,username,password FROM users--

This combines results from both queries, effectively retrieving usernames and passwords from the \texttt{users} table. It is particularly effective for quickly extracting sensitive information during exploratory phases.

By understanding these examples and implementing robust security practices such as parameterized queries, strict input validation, and minimal database permissions, developers can significantly reduce the risk of SQL injection attacks.

\section{Deeper Discussion on Contribution}
\label{Deeper Discussion}
Our work is specifically designed for the malicious code generation scenario, rather than as a general-purpose few-shot text generation framework. Our proposed methods and components are tailored to address the unique challenges of malicious code generation with limited training dataset. (1) Currently, few-shot text generation is often performed by providing a few input-output examples as prompts to large language models or by fine-tuning pre-trained models with a small amount of labeled data. In contrast, we employ a GAN-based framework to jointly train both the generator and discriminator, where the discriminator’s probability of judging the generated code as malicious serves as a reward signal to guide the training of the generator. Our method not only enables simultaneous malicious training of the generator and discriminator, but also allows the discriminator to provide more effective malcode-aware guidance to the generator, resulting in greater performance improvements compared to standard fine-tuning. (2) Moreover, other text generation methods cannot be directly applied to our scenario—on the one hand, limited training data may lead to poor quality in generated specific malicious code, and on the other hand, LLM models may refuse to generate the required malicious code in response to prompts related to malicious code generation due to safety reasons. Our approach has been experimentally validated to be both innovative and effective for malicious code generation in the security domain. 

\section{Related Work}
\textbf{SeqGAN} \cite{yu2017seqgan} is the first model that applied Generative Adversarial Networks (GANs) to sequence generation tasks, especially in text generation. SeqGAN utilizes reinforcement learning to optimize the generator, but it suffers from sparse rewards and unstable training, particularly when generating high-quality text. The sparse reward signal makes the generator's learning process slow and difficult, which limits the overall quality of the generated outputs.

\textbf{LeakGAN} \cite{guo2018long} introduces a leaky discriminator to address the gradient vanishing problem in SeqGAN. This stabilizes the training of the generator and accelerates learning. However, LeakyGAN still struggles with the diversity of the generated outputs and training instability. Despite providing more stable gradients, the generator still faces difficulties in generating more diverse and high-quality outputs.

\textbf{MaliGAN} \cite{che2017maximum} improves upon SeqGAN and LeakyGAN by using multiple discriminators to enhance the diversity of the generated sequences and address the mode collapse problem. While this improves the diversity of outputs, it increases the complexity of the model and training. Moreover, it still struggles with the instability of the training process and the monotony of the generated sequences.

\section{Methodology}

\subsection{Generator Loss in GANBERT}
\noindent The practical implementation computes mini-batch estimates as:

\begin{align}
    \mathcal{L}_{\text{adv}} &= -\frac{1}{N}\sum_{i=1}^N \log\left(1 - D_{\text{fake}}^{(i)}[k{+}1] + \epsilon\right), \\
    \mathcal{L}_{\text{feat}} &= \frac{1}{d}\left\|\frac{1}{N}\sum_{i=1}^N f_D(x_r^{(i)}) - \frac{1}{N}\sum_{j=1}^N f_D(x_g^{(j)})\right\|_2^2,
\end{align}

\noindent where $\epsilon$ ensures numerical stability, $N$ is the batch size, and $d$ is the feature dimension.

\section{Experiment Preparation}
\subsection{The Validity of Code Generation}
In the experiment, we present a comparative analysis of three versions of the Qwen-2.5coder model(1.5B)\citep{hui2024qwen2}: the base model without fine-tuning, the model fine-tuned with a specific amount of dataset, and the model trained using our proposed training framework. 
For the fine-tuning process, we utilized a subset of 1,000 samples from the SQLiDataset dataset \citep{solinsm_ml_sqli_xss_django_2023} obtained from the repository on GitHub. Additionally, we created 1,000 synthetic training samples based on the aforementioned dataset to augment our training data.

The fine-tuning was conducted using the architecture of Unsloth to train the Qwen-2.5coder model for two rounds.

After fine-tuning the model, we integrated it into the GANGRL-LLM framework. We then used the aforementioned 1,000 training samples to prompt the fine-tuned model for code generation. The generated code was subsequently fed into the discriminator for scoring. The scores were then used to guide the training process. This iterative process was repeated for 20 rounds, with each round generating 50 new code samples and performing corresponding scoring and model optimization.

\subsection{The Validity of Generated Dataset}
After fine-tuning with 1,000 samples and subsequently training the Qwen2.5coder model using the GANGRL-LLM framework, we utilized this enhanced model for code generation. To evaluate its performance, we created 2,000 prompts aimed at generating SQL injection (SQLi) code. The model-generated SQLi code was then used to train a detection model by performing substitution and expansion operations. This process was designed to verify whether our generated malicious code data could enhance the detection capabilities of the model.
We utilized the publicly available SQLi dataset \citep{kaggle_sql_injection_dataset}, using 4,000 samples for training (2,000 SQLi samples and 2,000 benign samples) and 1,000 samples for testing (800 SQLi samples and 200 benign samples) to simulate the detection capability of the model with a limited number of samples. We generated additional data using the Qwen2.5coder model, which was fine-tuned on 1,000 samples and subsequently trained using the GANGRL-LLM framework.
\begin{table}[h]
\centering
\small
\begin{tabular}{c|c}
\toprule
\textbf{Model} & \textbf{Score} \\
\midrule\midrule
 Qwen2.5coder & 7.11 \\

 Finetuning with 1000 & 7.02 \\

 GANGRL-LLM & \textbf{7.136} \\

 RL & 6.558 \\
\bottomrule
\end{tabular}
\caption{Scores for different models with varying numbers of fine-tuning datasets.}
\label{tab:scoresrl}
\end{table}

\begin{table*}[h]
\centering
\small
\begin{tabular}{p{0.25\textwidth}|p{0.65\textwidth}}
\toprule
\multicolumn{2}{c}{\textbf{Prompt}} \\
\midrule\midrule
\multicolumn{2}{p{0.9\textwidth}}{Please help me generate a SQL injection code. The SQL injection vulnerability in the provided code utilizes the \colorbox{LightPink}{regexp\_substring} function, exploits the \colorbox{LightPink}{char} function for potentially crafting a payload, and depends on improper handling of \colorbox{LightPink}{NULL values and logical conditions} to mask the malicious input.} \\
\toprule
\textbf{Model} & \textbf{Generated Code}\\
\midrule\midrule
RL & 1"   )    )     )   and char  (  113  )  =  char  (  regexp\_substring  (  char  (  113  )  ||char  (  113  )  ||char  (  113  )  ||char  (  113  )  ,  '.*?'  ,  1  )  ,  1  )   and   (  char  (  113  )  =  char  (  regexp\_substring  (  char  (  113  )  ||char  (  113  )  ||char  (  113  )  ||char  (  113  )  ,  '.*?'  ,  1  )  ,  1  ) \\
\bottomrule
\end{tabular}
\caption{Examples of generated SQL code from RL training models.}
\label{tab:rl}
\end{table*}

\begin{table*}[ht]
\centering
\small
\resizebox{\textwidth}{!}{
\begin{tabular}{l|l|l|l|l|l}
\toprule
\textbf{Operation} & \textbf{MODEL NAME} & \textbf{Accuracy} & \textbf{Precision} & \textbf{Recall} & \textbf{F1} \\ \midrule\midrule
\multirow{5}{*}{no operation} & CNN          & 0.904 & 0.9985835694050992 & 0.88125  & 0.93625498 \\ 
                              & Naive Bayes  & 0.828 & 0.8243801652895262 & 0.9975   & 0.902714932 \\
                              & SVM          & 0.489 & 1 & 0.36125   & 0.530762167 \\ 
                              & KNN         & 0.807 & 0.8056394763343404 & 0.8975   & 0.892359175 \\

                              & Decision Tree & 0.917 & 0.9986091794158554 & 0.8975   & 0.945358789 \\ \midrule
\multirow{5}{*}{Switch 1000 training datasets} & CNN & \textbf{0.91}  & \textbf{1}                   & \textbf{0.8875}   & \textbf{0.940397351} \\ 
                                               & Naive Bayes  & \textbf{0.829} & \textbf{0.8259067357512954} & 0.99625& \textbf{0.903116147} \\ 
                                               & SVM          & \textbf{0.49} & 1                   & \textbf{0.3625}   & \textbf{0.532110092}  \\ 
                                               & KNN          &0.806 & 0.8048289738430584 & 1  & 0.891861761 \\ 
                                               & Decision Tree & \textbf{0.92} & \textbf{1}                   & \textbf{0.9}  & \textbf{0.947368421}  \\ \midrule
\multirow{5}{*}{Add 1000 training datasets}   & CNN          &\textbf{0.927} & \textbf{1} & \textbf{0.90875}   & \textbf{0.952193844} \\ 
                                               & Naive Bayes  & \textbf{0.83} & \textbf{0.8247422680412371} & \textbf{1}   & \textbf{0.903954802} \\ 
                                               & SVM          & \textbf{0.722} & 1 & \textbf{0.6525}
 & \textbf{0.789712557} \\ 
                                               & KNN          & \textbf{0.817} & \textbf{0.8138351983723296} & 1  & \textbf{0.897363993} \\ 
                                               & Decision Tree & \textbf{0.936} & \textbf{1}                   & \textbf{0.92}  & \textbf{0.958333333} \\ \midrule
\multirow{5}{*}{Add 2000 training datasets}   & CNN          & \textbf{0.928}  & \textbf{1} & \textbf{0.91}  &\textbf{0.952879581} \\ 
                                               & Naive Bayes  & \textbf{0.83}  & \textbf{0.8247422680412371}  & \textbf{1}   & \textbf{0.903954802} \\ 
                                               & SVM          & \textbf{0.722} & 1 & \textbf{0.6525}& \textbf{0.789712557} \\
                                               & KNN          & \textbf{0.817} &\textbf{ 0.8138351983723296} & 1        & \textbf{0.897363993} \\ 
                                               & Decision Tree & \textbf{0.936}  & \textbf{1}                   & \textbf{0.92}  & \textbf{0.958333333} \\ \bottomrule
\end{tabular}
}
\caption{Models performance for different training datasets and operations generated by Llama3.2}
\label{tab:ml2}
\end{table*}

\begin{table*}[ht]
\centering
\small
\resizebox{\textwidth}{!}{
\begin{tabular}{l|l|r|r|r|r}
\toprule
\textbf{Operation} & \textbf{MODEL NAME} & \textbf{Accuracy} & \textbf{Precision} & \textbf{Recall} & \textbf{F1} \\ 
\midrule\midrule
\multirow{5}{*}{No operation (1000)} & CNN          & 0.442 & 0.7079 & 0.515 & 0.596 \\ 
                                    & Naive Bayes  & 0.810 & 0.8138 & 0.9888 & 0.893 \\
                                    & SVM          & 0.325 & 1.0000 & 0.156 & 0.270 \\ 
                                    & KNN          & 0.794 & 0.7988 & 0.9925 & 0.852 \\
                                    & Decision Tree & 0.814 & 0.8120 & 0.9988 & 0.896 \\ 
\midrule
\multirow{5}{*}{Add 500}            & CNN          & 0.795 & 0.8195 & 0.954 & 0.882 \\ 
                                    & Naive Bayes  & 0.818 & 0.8153 & 0.9988 & 0.898 \\ 
                                    & SVM          & 0.800 & 0.8000 & 1.0000 & 0.889 \\ 
                                    & KNN          & 0.800 & 0.8000 & 1.0000 & 0.889 \\ 
                                    & Decision Tree & 0.814 & 0.8120 & 0.9988 & 0.896 \\ 
\midrule
\multirow{5}{*}{Add 1000}           & CNN          & 0.904 & 0.9944 & 0.885 & 0.937 \\ 
                                    & Naive Bayes  & 0.915 & 0.9945 & 0.8988 & 0.944 \\ 
                                    & SVM          & 0.800 & 0.8000 & 1.0000 & 0.889 \\ 
                                    & KNN          & 0.800 & 0.8000 & 1.0000 & 0.889 \\ 
                                    & Decision Tree & 0.902 & 1.0000 & 0.878 & 0.935 \\ 
\bottomrule
\end{tabular}
}
\caption{Model performance trained by less samples for different training datasets and operations}
\label{tab:ml3}
\end{table*}

\section{Experiment Details}
\subsection{Scoring Basis}
\label{Scoring Basis}
The evaluation criteria includes:
\textbf{Adherence to the Prompt}: Measuring how closely the generated code aligns with the requirements specified in the prompt.
\textbf{Complexity of the Code}: Assessing the intricacy and sophistication of the generated SQL injection code.
\textbf{Effectiveness of the SQL Injection}: Evaluating whether the generated SQL injection code can successfully exploit vulnerabilities in target systems.
\textbf{Correctness of the Code}: Verifying the syntactic and logical accuracy of the generated code.

Each criterion  is quantitatively scored using predefined standards established by the Qwen2.5Turbo model. The overall score (ranging from 1 to 10) provides a holistic view of the models' performance in generating high-quality malicious code. By utilizing Qwen-Turbo, we ensure that our evaluations are both rigorous and consistent, leveraging the state-of-the-art capabilities of Qwen2.5Turbo to deliver reliable and actionable insights.

\subsection{Lab Proc}
\begin{figure}[!ht]
  \centering
  \includegraphics[trim=80 80 20 40, clip, width=1.1\columnwidth, keepaspectratio]{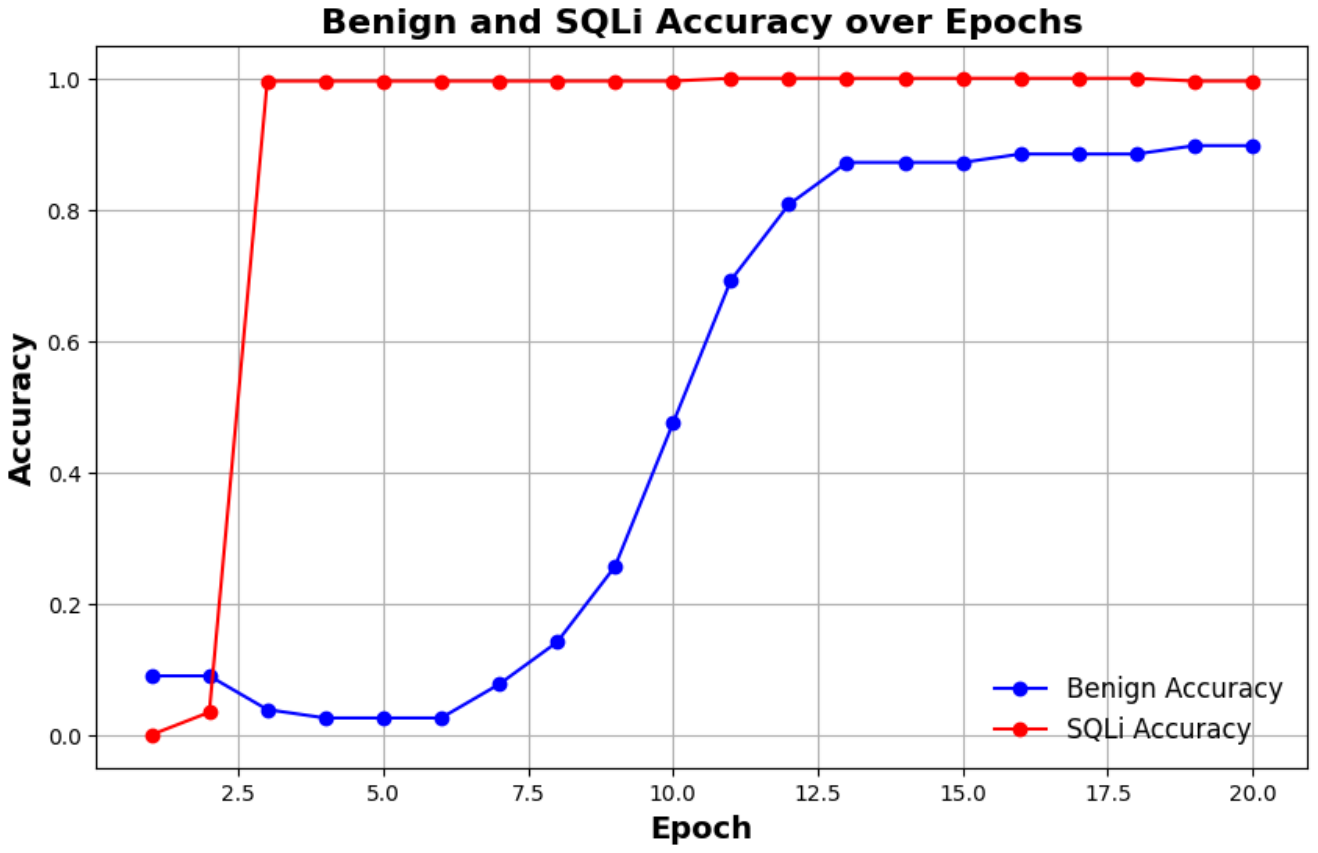}
  \caption{Accuracy in training process.}
  \label{figure:accuracy}
\end{figure}
During the training process, the change in accuracy with the variation of training epochs is shown in the Figure~\ref{figure:accuracy}. As we can see, even though there are few labeled samples, the model's detection capability gradually improves with the increase in training epochs.

\section{Experiment with RL}
During the design of our optimization strategy, we initially employed policy gradient optimization from reinforcement learning(RL). 
\[
\nabla_{\theta} J(\theta) = \mathbb{E}_{\pi_{\theta}}\left[ \sum_{t=0}^{T} \nabla_{\theta} \log \pi_{\theta}(s_t, a_t) \cdot R_t \right].
\]

Where:

\begin{itemize}
  \item $\pi_{\theta}(s_t, a_t)$ is the policy probability for taking action $a_t$ at state $s_t$.
  \item $R_t$ is the return starting from time step $t$, typically the cumulative reward from time step $t$ to the end.
  \item $\nabla_{\theta} \log \pi_{\theta}(s_t, a_t)$ is the gradient of the log of the policy function with respect to $\theta$, representing how sensitive the policy is to changes in the parameters at state $s_t$ and action $a_t$.
\end{itemize}
However, after training with a limited number of samples, we found that using only the discriminator's output probability of maliciousness as a reward to guide the generator's optimization led to suboptimal results. Specifically, while the generated code was indeed malicious, it often deviated from the methods and requirements specified by the prompt, resulting in poor generation quality according to Table~\ref{tab:scoresrl}.

Therefore, we employed a generated reward mechanism combined with the cross-entropy loss from supervised learning to ensure that the generated code adheres to the requirements of the original code. It has been demonstrated that this approach also achieves higher quality scores.

In Table~\ref{tab:rl}, it shows the code generated by the model training with RL method.Although this piece of code exhibits some of the characteristics described, particularly in utilizing regexp\_substring and char functions to construct SQL injection payloads, it does not fully align with the description in terms of handling NULL values and employing complex logic to bypass security measures. Specifically, the code lacks evident application of techniques for handling NULL values and implementing complex logic bypasses. As a result, it does not completely match the vulnerability characteristics outlined in the description.
\subsection{Malcode analysis}
\label{malcode analysis}
In the first SQL injection code where char(3702) might produce an invisible or invalid character. This makes the attack more likely to be blocked by protective systems. It lacks complex conditions or logic bypass mechanisms, making it relatively straightforward but potentially ineffective against robust defenses.

The third SQL injection code also uses character concatenation to form a malicious payload, combining regexp\_substring with the char function. The use of "--" at the end serves to comment out the rest of the query, further masking the SQL injection attempt. However, this code segment is relatively simple compared to the second one. It lacks advanced logical conditions or sophisticated evasion techniques, making it less effective in terms of complexity and ability to bypass security measures.

In terms of conformity to the prompt and complexity and effectiveness of the SQL injection, the second SQL injection code is the most aligned and sophisticated. It closely matches the prompt by utilizing both the regexp\_substring and char functions to craft a complex payload, incorporating multiple logical conditions and character concatenations. The inclusion of "zjzq" = "zjzq" adds an additional layer of obfuscation, making the injection harder to detect and more likely to bypass security measures. This indicates that our training framework can produce higher-quality code.
\section{Llama Model Experiment}
We also conducted experiments using the LLama 3.2 (1B) model as the baseline. Under identical conditions, including the same dataset and data processing methods, we generated 2000 samples on the same test set. Subsequently, we replaced the corresponding samples in the dataset with these generated ones, which led to the final experimental results summarized in the Table~\ref{tab:ml2}.

In the vast majority of cases, the code generated by our model enhances the detection capability of the original model. The only exception is the KNN model, which shows a slight decrease in performance after data replacement. Our analysis indicates that some of the generated code is more complex and requires benign-looking code features to bypass detection. Since the KNN model relies on nearest neighbor labels for judgment, it tends to misclassify these complex samples as benign. As the number of complex samples increases, the trained code also improves the overall detection capability of the model.

\end{document}